\documentclass[12pt,a4paper]{article}
\pdfoutput=1
\usepackage[format=hang]{caption}
\usepackage{epsfig,amssymb,amsmath,graphicx,subcaption,xcolor,sidecap,float,comment,verbatim, ulem,hyperref}
\newcommand*\rfrac[2]{{}^{#1}\!/_{#2}}
\newcommand{\be}{\begin{equation}}
\newcommand{\ee}{\end{equation}}

\DeclareMathOperator{\Tr}{Tr}
\DeclareMathOperator{\tr}{tr}

\def\d{\delta}

\begin{document}

\numberwithin{equation}{section}
{
\begin{titlepage}
        \begin{flushright}
                {\normalsize YITP-16-93} \\
        \end{flushright}
\begin{center}

\hfill \\
\hfill \\

{\Large \bf Inspecting non-perturbative contributions to the Entanglement Entropy via wavefunctions  }\\

\vskip 0.4in

{\bf Arpan Bhattacharyya${}^{a,d}$, Ling-Yan Hung${}^{a,b,c}$, P.H.C. Lau${}^{e}$ and Si-Nong Liu${}^{a}$  }

\vskip 0.2in

{\it ${}^{a}$ Department of Physics and Center for Field Theory and Particle Physics,\\ Fudan University,
220 Handan Road, 200433 Shanghai, China} \vskip .5mm
{\it ${}^{b}$State Key Laboratory of Surface Physics and Department of Physics,\\ Fudan University,
220 Handan Road, 200433 Shanghai, China}

{\it ${}^{c}$ Collaborative Innovation Center of Advanced  Microstructures,\\
Nanjing University, Nanjing, 210093, China.}\vskip.5mm
{\it ${}^{d}$ Centre For High Energy Physics, Indian Institute of Science, 560012 Bangalore, India}\vskip.5mm
{\it${}^{e}$ Yukawa Institute for Theoretical Physics, Kyoto University, Kitashirakawa Oiwakecho \\
Sakyo-ku, Kyoto City, Kyoto 606-8502, Japan}

\end{center}

\vskip 0.25in

\begin{center} {\bf ABSTRACT } 
\end{center}
In this paper, we would like to systematically explore the implications of non-perturbative effects on entanglement in a many body system. Instead of pursuing the usual path-integral method in a singular space, we attempt to study the wavefunctions in detail. We begin with a toy model of multiple particles whose interaction potential admits multiple minima. We study the entanglement of the true ground state after taking the tunnelling effects into account and find some simple patterns. Notably, in the case of multiple particle interactions, entanglement entropy generically decreases with increasing number of minima. The knowledge of the subsystem actually increases with the number of minima. The reduced density matrix can also be seen to have close connections with graph spectra.
In a more careful study of the two-well tunnelling system, we also extract the exponentially suppressed tail contribution, the analogues of instantons.  
To understand the effects of multiple minima in a field theory, it inspires us to inspect wavefunctions in a toy model of bosonic field describing quasi-particles of two different condensates related by Bogoliubov transformations. We find that the area law is naturally preserved. This is probably a useful set of perspectives that promise wider applications.

\vfill

\noindent \today

\end{titlepage}
}

\newpage

\tableofcontents


 




\section{Introduction}

The study of many body entanglement has given us insights into new phases of matter, leading to a classification of topological orders beyond the Landau paradigm \cite{ Levin, Preskill, Wen}. Entanglement entropy has also breathed new life into the study of quantum gravity. Efforts to understand the Ryu-Takayanagi formula \cite{Ryu,Ryu1} that computes entanglement entropy  in the AdS/CFT correspondence has led to novel perspectives that connect the duality to tensor-networks \cite{Swingle, TensorNetwork,TensorNetwork2}. The latter surprisingly captures many of the salient features of the correspondence. In high energy physics, the default framework has been quantum field theory. Given the importance in understanding the entanglement pattern of a many-body system\footnote{Interested readers are referred to some of these references \cite{Manybody,Manybody1,Manybody2,Manybody3,Manybody4,Manybody5,Manybody6,Manybody7,Manybody8,Manybody9,Manybody10,Manybody11,Manybody12}. The list is by no means complete and interested readers are encouraged to refer to the references and citations of these papers.},  it is thus of fundamental significance to understand what kind of entanglement can be supported by quantum field theory in principle. 

As the most broad-stroke qualitative question, it is natural to ask whether the entanglement entropy of the ground state of a local field theory always satisfies an area law. There is overwhelming evidence that this is indeed the case, from both numerical studies and exact calculations in field theories, such as CFTs in 1+1 dimensions, free theories and large N theories \cite{FT,FT1,FT2,FT3,FT3,FT4} in which the calculation becomes manageable. There are also perturbative calculations in theories such as $\phi^4$ theory \cite{FT5,FT6} that provide some insights into the correction to the entanglement entropies from some weak interactions \cite{FT6a}. On the other hand, we have the AdS/CFT correspondence that allows one to probe into the strongly coupled regime of the field theory \cite{Ryu1,FT7}.

There are various aspects that still remain unsatisfactory. For one, the corrections included in these computations are predominantly perturbative. Also, the entanglement entropy is a rather crude probe of entanglement. It misses many minute details of the actual pattern of entanglement that could only be explored with a better control of the entanglement spectrum, such as in the classic example of topological orders \cite{edgebulk1}, and more recently, in identifying chaotic states \cite{Alioscia}. Moreover, the entanglement entropy has to be computed via the replica trick \cite{FT,FT1},  which, given its technical difficulty, easily masks the physics. After all, the calculation does not directly provide insights of entanglement between physical degrees of freedom. Rather, it relies on heavy machinery that introduces singularities into a path-integral. The procedure of regulating subsequent divergences can potentially mask the true underlying physics.

In this paper, we are motivated by the intention of studying non-perturbative contributions to the entanglement entropy, and understand, at least in some special contexts, what role they play in the entanglement of the ground state.  One of the most important non-perturbative corrections to field theory comes from instantons, which are extra (topologically) non-trivial saddles in the (Euclidean) path-integral. However, it is not entirely obvious how they alter the entanglement structure of the ground state. To address the question, we follow here the more intuitive path of focusing on the wavefunction. It is well understood that instantons describe tunnelling events between local minima, and that we are only capable of employing perturbation theory systematically around each of these minima.  Instantons have been explored in a wide range of theories,  from $\phi^4$ theories with two minima to Abelian and non-Abelian gauge theories such as the QCD \cite{Inst1,Inst2,Inst3} with essentially infinitely many minima. In the case of the $\phi^4$ theory (in a compact space), instantons describe tunnelling events between the two minima \cite{Inst1}. In QCD, the study of tunnelling allows one to identify the $\theta$-vacua \cite{Inst2,Inst3}. The true ground state is in fact a linear combination of the naive vacua obtained by perturbing around one of the minima of the potential. 

Motivated by these considerations, we begin with a study of a simple quantum mechanical problem, in which multiple particles enjoy an interaction potential that has multiple minima. The true ground state is a linear combination of states where each one is a peak around one minimum. In these cases, the reduced density matrix can be obtained explicitly, giving a complete picture of the entanglement structure of the true ground state. Along the way, we find an amusing connection between the spectra of the reduced density matrices and the Cartan matrices of the $A_n$ Lie groups. The relation between these is part of a wider connection between the entanglement spectrum and graph spectra \cite{Graph, Graph1,Graph2}, where the adjacency matrix of some graphs describes hopping between different states in the reduced density matrix.  

Then we move onto field theory. To borrow the insights obtained from the quantum mechanical problem, we start with the true ground states of a theory with multiple minima after taking tunnelling into account. Instead of working directly with instantons, inspired by the true ground states of the quantum mechanical system, we play the same game of writing down linear combinations of states, each of which corresponding to a wavefunction peaked around one of the minima. To do so, we select a reference basis state defined perturbatively around one minimum. This is a complete set of basis states where the perturbative vacuum defined around another minimum can be related to the reference set of states by a Bogoliubov transformation \cite{Bogo,Bogo1,Bogo1a, Bogo2, Bogo3}. We will demonstrate these explicitly using a set of free bosons in 1+1 dimensions as an example. For $\phi^4$ theories, it is natural to work with bosonic excitations around different minima, each of them corresponds to a superconducting condensate \cite{Bogo2,Bogo3}. Using this method, one could explicitly see that the area law is expected to be preserved, even with instanton corrections, as also observed using direct field theoretic computation employing the replica trick \cite{me}.  

The main lesson we have learnt in these attempts is that the condensed matter folklore -- that the ground states of local theories respect the area law indeed remain true in each of the examples even as we include non-perturbative corrections.

\section{Toy example of a 2-particle system \label{sec:2_par}}

Let us begin with a warm-up exercise studying the entanglement between 2 particles, located at $x_1$ and $x_2$, interacting via a potential $V(x_1-x_2)$. The 2-particle time-independent Schr\"odinger equation is given by \cite{qm}
\be
H \left| \Psi_n(x_1,x_2) \right\rangle = E_n \left| \Psi_n(x_1,x_2) \right\rangle,
\ee
with the hamiltonian
\be
H= -\frac{1}{2m_1}\partial_{x_1}^2 - \frac{1}{2m_2}\partial_{x_2}^2 + V(x_1-x_2).
\ee
To illustrate the idea, we will take $V(x)$ to be a function with 2 minima located at $x=x_1-x_2=0$ and $x=a$. The center of mass behaves like a free particle. Suppose the potential is extremely narrow and deep near $x=0$, $a$, and that $V(0)=V(a)=0$, then the ground state in the center of mass frame is simply given by a wavefunction that is an equal weighted sum of delta functions peaking at $x=0$ and $x= a$.
The low lying eigenfunctions take the form,
\be
\left| \Psi_k(x_1,x_2) \right\rangle = \frac{\mathcal{N}}{\sqrt{2}}\sum_{x_1,x_2} e^{i k. X} \left( \left| x=0 \right\rangle + \left| x=a \right\rangle  \right), \qquad X= \frac{m_1 x_1+ m_2 x_2}{(m_1+m_2)}
\label{eqn:2part}
\ee
where $\mathcal{N}$ is the appropriate normalisation and the true ground state is located $k=0$. $x_1,x_2$ are discretised such that the integrations over $x_1$ and $x_2$ are replaced by summations. $X$ is the centre of mass coordinate and $k.X$ denotes the usual scalar product between two vectors.

The density matrix constructed from $|\Psi_k\rangle$ is given by $\rho = \left|\Psi_k \right\rangle \left\langle \Psi_k \right|$,
and the reduced density matrix upon integrating out particle 2 is given by
\be
\rho_1 = \mathcal{N}^2\sum_{x_1} \left( \left|x_1 \right\rangle \left\langle x_1 \right| + \frac{1}{2}(e^{ik \frac{m_1 a}{m_1+m_2}} \left| x_1 \right \rangle \left\langle x_1+a \right| + e^{- ik \frac{m_1 a}{m_1+m_2}} \left|x_1 \right\rangle \left\langle x_1-a \right| )\right) \,, \label{eqn:red_den_2par}
\ee
which explicitly has unit trace. 

The off-diagonal terms are essentially hopping terms. Note that for a potential $V(x)$ with a single minimum, the reduced density matrix corresponds to a maximally entangled state where we have no knowledge of particle 2.

Here, we would like to pause and note that at $k=0$, this reduced density matrix is precisely proportional to the Cartan matrix of $A_N$, where $N$ is now given by the total number of sites of the discretised system. 
The Cartan matrix of $A_N$ is equal to $2 \mathbb{I} - R$, where $R$ is the adjacency matrix of the Dynkin diagram of $A_N$ defined by \footnote{The adjacency matrix  $R$ of a graph is defined such that a matrix element $R_{ij}$ counts the number of links connecting the two nodes labelled $i$ and $j$. For a best beginner's review of graph spectra, see for example Wikipedia \cite{link} and references therein. }
\be
R_{ij} = 1, \qquad i = j\pm1.
\ee

It is not surprising that the reduced density matrix is connected to graphs. The Cartan matrix is connected to positive definite matrices, which alone is a good reason why it should show up as reduced density matrices. We leave this important and fun topic for future work.

To compute the entanglement entropy, we use the replica trick and compute $\Tr \rho_1^n$.
This is a combinatorics problem asking for the number of ways to hop $n$ steps and returning to the starting point. The number of steps in $+a$ must equal the number of $-a$ steps. Suppose they are $m$, then we are taking $(n-2m)$ steps hopping on the same spot. The contribution to $\Tr \rho_1^n$ from such a term is
\be
f_n(m)= \mathcal{N}^{2n}\frac{n!}{m!m!(n-2m)!}\frac{1}{2^{2m}} \, .
\ee
Then altogether we have
\be
\Tr \rho_1^n =\frac{1}{\mathcal{N}^2} \sum_{m=0}^{\left|\rfrac{n}{2}\right|} f_n(m) \,.
\ee
This is evaluated formally and gives
\be
\Tr \rho_{1}^{n}= \mathcal{N}^{2(n-1)}\Big(\frac{2^{n}\Gamma(n+\frac{1}{2})}{\sqrt{\pi}\Gamma(n+1)}\Big).
\ee
The entanglement entropy for two minima is
\be
S^{2}_{EE}=\lim_{n\rightarrow 1} \frac{\ln ( \Tr \rho_{1}^{n})}{1-n}= - 2\ln \left(\mathcal{N} \right) +1 -\ln(2) -\gamma - \psi ^{(0)}\left(\frac{3}{2} \right),
\ee
where $\gamma$ and $\psi ^{(0)}$ denotes the Euler's constant and the digamma function, respectively.  \\ Now we compare it with the case when there is only one minimum
\be
\rho_{1}= \mathcal{N}^2 \sum_{x_1}\left|x_{1} \right\rangle \left\langle x_{1} \right| \,.
\ee
The entanglement entropy for a single minimum is
\be
S^{1}_{EE}= -2 \ln (\mathcal{N}) .\,
\ee
So the change in entanglement entropy between this two cases is
\be
\Delta S_{EE}=S^{1}_{EE}-S^{2}_{EE}=0.306853 \, .
\ee
There is a reduction of entanglement. To elaborate further, in the case where there is exactly 1 minimum, it means that the position of 1 is completely locked with 2. When 2 is not measured, the position of 1 is completely undetermined, with the same probability distributed over all possible sites. This is the infinite dimensional analogue of the bell state $\frac{1}{\sqrt{2}} \left( \left| \uparrow \uparrow \right\rangle  + \left| \downarrow\downarrow \right\rangle \right)$ when the two spins are maximally entangled. It leads to a maximal amount of ignorance when particle 2 is traced out. Now one can see that any departure from this pattern of entanglement would always reduce the entanglement entropy. Or in other words, in the case of two minima, when particle 2 is located at some position, particle 1 can be found in two different locations. This means that particle 1 has some ``internal'' distribution that does not depend on what we do with 2, and so there could be information gained from measuring particle 1 even without particle 2, unlike the single minimum case when their behaviour are completely locked.

In the next section, we will generalise to the case of arbitrary number of minima. 
\subsection{Generalisation to multiple minima}

The previous calculation can be generalised to the case where the potential $V(x)$ has $N$ minima, located at $\{L_i\}$, $1 \leq i\leq N.$
The reduced density matrix is
\be
\rho_1(N)= \mathcal{N}^2 \sum_{x_1}\left(\left|x_1\right\rangle \left\langle x_1 \right| + \frac{1}{N}\left(\sum_{i\neq j}e^{i k (L_i-L_j)} \left|x_1 \right\rangle \left\langle x_1+(L_i-L_j) \right| \right) \right) \,.
\ee
Now we apply the replica trick and compute $\Tr\rho^n_1$.  For simplicity, and in fact generically, $L_{ij}= L_i-L_j$ are not multiples of each other, and there are thus $N(N-1)/2$ distinct hopping step sizes.  
\be
n_0+ \sum_{i,j} n_{ij} =n \,,
\ee 

\be
\sum_{i,j}n_{ij}(L_i-L_j) =0 \,,
\ee where $n_0$ is the number of standstill hops. 

The contribution from each such set of closed paths is given by
\be
f_n(\{n_{ij}\}) = \mathcal{N}^{2n}  \frac{n!}{N^{n-n_0} \, n_0!\, \prod_{i,j} (n_{ij}!)} \,.
\ee

For simplicity, let us compute the case of $n=2$ which gives us the tightest lower bound on the entanglement entropy. 
For $n=2$, the only allowed hopping is to take either $n_{ij}=n_{ji}=1,\, n_0=0$, or $n_0=2$ which gives
\be
\Tr \rho_1^2 = \mathcal{N}^2 \left(1+ \frac{N(N-1)}{N^2} \right)= \mathcal{N}^2 \left(2 -\frac{1}{N}\right).  
\ee
We note that the entanglement entropy must be a monotonically decreasing function of $N$, the number of minima. It follows from the fact that with more minima, we actually gain knowledge of the first particle, for the same reason discussed previously,  that two minima potential reduces entanglement compared to the case of single minimum. 
\subsection{Generalisation to many particles with two minima }
\begin{figure}[ht]
        \centering
        \includegraphics[width=6cm]{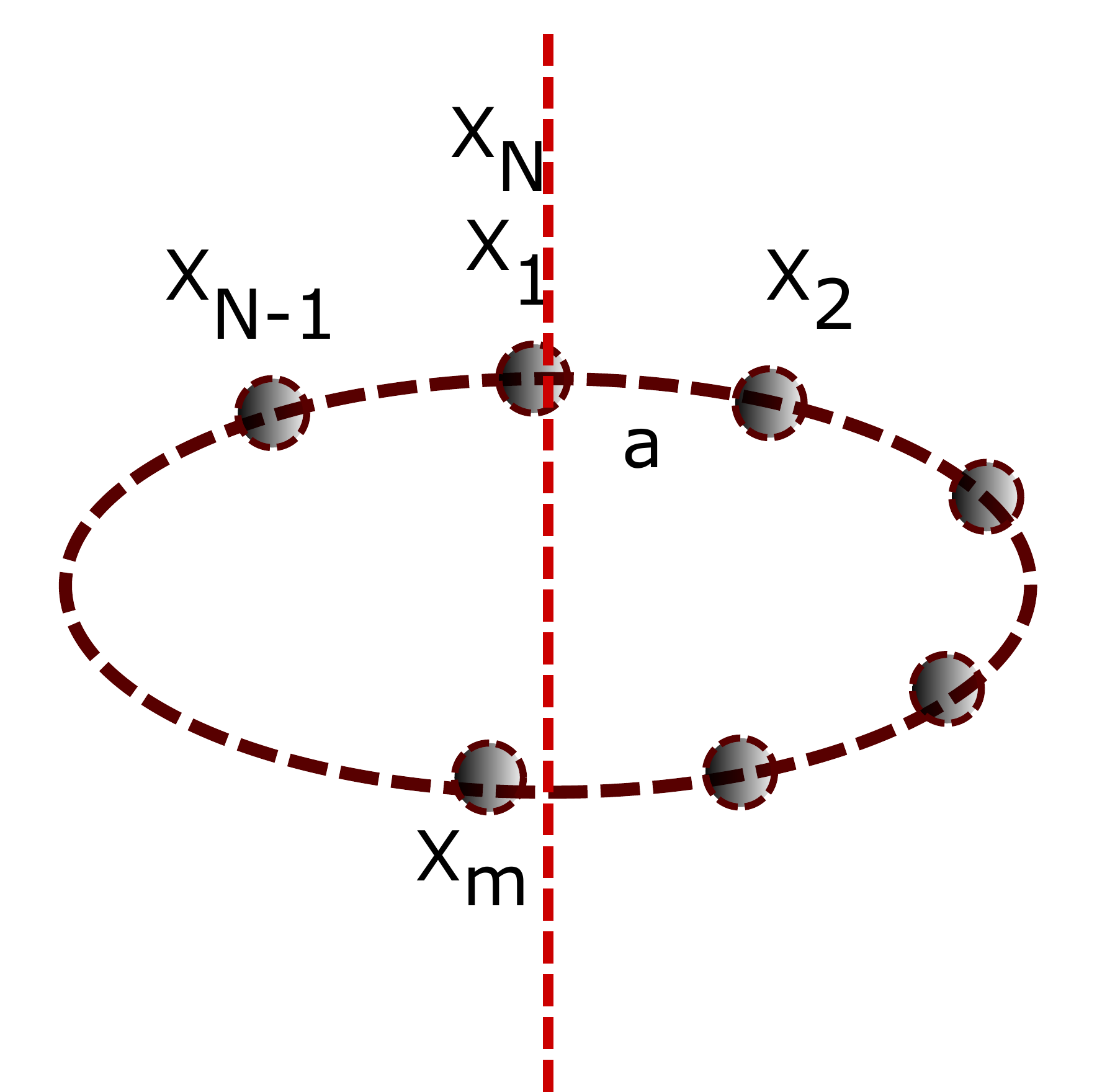}
        \caption{$N$ particles on a ring. $x_1, x_{2}\cdots x_{N}$ denote the positions of these particles. $a$ denotes the separation between any two particles. Dashed line in red denotes that the first m particles are traced out when we compute the reduced density matrix. }
        \label{Fig:ring}
\end{figure}

\subsubsection*{Particles on a ring: periodic boundary condition}
In this section we generalise the previous calculation to $N$ particles with the same form of potential $V(x_{i+1},x_i)$ admitting only two minima between the nearest neighbours as in the first part of Section \ref{sec:2_par}. For simplicity we set $k=0$ here.  We consider a set-up where these $N$ particles are placed on a ring, i.e. with a periodic boundary condition. The wavefunction is given by
\be
\left|\psi(x_1, x_2,\cdots x_{N})\right\rangle = \frac{\mathcal{N}}{\sqrt{N!}}\sum_{\Delta_{ij}} \Big(\left|x_{1},\Delta_{ij}=0 \right\rangle + \left| x_{1}, \Delta_{ij}=a,-a \right\rangle \Big) \,,
\ee
where, $i,j \geq 1.$ $\Delta_{ij}$ denotes the separation between the particles $i$ and $j$ and $\sum_{x_1} \left|x_{1}, \Delta_{ij}=a,-a\right\rangle$ denotes all the configurations when any one of the  $\Delta_{ij}=x_i-x_j$ are non zero and equal to $a.$ The periodic boundary condition implies $\sum_{i,j}\Delta_{ij}=0$ which gives a very tight constraint. Since $\Delta_{ij}$ $\forall \, i,j$ can only take values either $a$ or $0$, this constraint implies either all $\Delta_{ij}=0$ or only one of them can be non-vanishing and take value $\Delta_{ij}=a$ when $i < j\,\, \forall \, i,j =\, 1 \cdots N$. The latter automatically forces $\Delta_{x_{N},x_{1}}=-a$. We further simplify the form of the wavefunction by taking $x_1$ as the reference point
\begin{align}
\begin{split}
|\psi>=\frac{\mathcal{N}}{\sqrt{N!}}\sum_{\Delta_{ij}} &\Big(\left|\psi_{1}(x_1,\Delta_{ij}) \right\rangle + \left|\psi_{2}(x_1,\Delta_{ij})\right\rangle+ \left|\psi_{3}(x_1,\Delta_{ij}) \right\rangle+ \left|\psi_{4}(x_1,\Delta_{ij}) \right\rangle \Big) \, ,
\end{split}
\end{align}
where
\begin{align}
\begin{split} \nonumber
\left|\psi_{1}(x_1,\Delta_{ij})\right\rangle=&\left|x_1,\Delta_{ij}=0\,\,\forall i,j \leq x_m\right\rangle \otimes\\ &  \left|x_1,\Delta_{x_{m},x_{m+1}}=0,\Delta_{x_{N},x_1}=0,\,\, \Delta_{ij}=0\,\,\forall \, i,j = x_{m+1}\cdots x_{N}\right\rangle \,,\\
\left|\psi_{2}(x_1,\Delta_{ij})\right\rangle = &\left|x_1,\Delta_{ij}=0\,\,\forall \, i,j \leq x_m\right\rangle \otimes\,\\& \left|x_1,\Delta_{x_{m},x_{m+1}}=a,\Delta_{x_{N},x_{1}}=-a,\Delta_{ij}=0\,\,\forall \, i,j = x_
{m+1}\cdots x_{N}\right\rangle \, ,\\
\left|\psi_{3}(x_1,\Delta_{ij})\right\rangle=&\left|x_1,\Delta_{ij}=0\,\,\forall \,i,j \leq x_m\right\rangle \otimes \\& \left|x_1,\Delta_{x_{m,m+1}}=0,\Delta_{x_{N},x_{1}}=-a,\Delta_{ij}=a\,\,\forall \, i,j = x_{m+1}\cdots x_{N}\right\rangle \,, \\
\left|\psi_{4}(x_1,\Delta_{ij})\right\rangle=&\left|x_1,\Delta_{ij}=a\,\,\forall \, i,j \leq x_m\right\rangle \otimes \\& \left|x_1,\Delta_{x_{N},x_1}=-a,\,\Delta_{x_{i},x_{j}}=0\,\,\forall \, i,j= x_{m}\cdots x_{N}\right\rangle \,.\\
\end{split}
\end{align}
Here, the total wavefunction is expressed as a tensor product of states living in the two regions. The periodic constraint then leads to four possible configurations of the states which are labelled as ($\psi_1,\psi_2,\psi_3,\psi_4$). $\psi_1$ corresponds to state with $\Delta_{ij}=0$ for all sites while $(\psi_2,\psi_3,\psi_4)$ correspond to states with $\Delta_{ij}\neq0$ at the boundary $i=x_m,j=x_{m+1}$, at region $i,j\leq x_m$,  and at region $x_m < i,j < x_N$, respectively. 
 Now the reduced density matrix can be computed by integrating out the first $m$ particles
\begin{align}
\begin{split}
\rho_{A}=\frac{\mathcal{N}^2}{N!}\Big(&(m-1)|x_{m}=x_1+a,\Delta_{x_{m} x_{m+1}}=0,\Delta_{x_{N} x_1}=-a> <x_{m}=x_1+a,\Delta_{x_{m} x_{m+1}}=0,\\& \Delta_{x_{N} x_{1}}=-a |+[|x_{m}=x_1,\Delta_{x_{m} x_{m+1}}=0,\Delta_{x_{N} x_{1}}=0>+|x_{m}=x_1+a,\Delta_{x_{m} x_{m+1}}=0,\\& \Delta_{x_{N} x_{1}}=-a>+|x_{m}=x_1,\Delta_{x_{m} x_{m+1}}=a, \Delta_{x_{N}, x_{1}}=-a>][\text{c.c}]\Big),
\end{split}
\end{align}
where c.c denotes the complex conjugate. 
Up to an overall normalisation factor
\be
\Tr\rho_{A}^2= 2[(m-1)^2+ 2 (m-1)+ \frac{N^2}{2}-N(m-1)+1].
\label{eqn:renyi2}
\ee

Finally the second R$\acute{\text{e}}$nyi entropy is
\be
S_2=-\ln (\Tr \rho_{A}^{2}) \,.
\ee
Note that we have used the fact that there are $(m-1)$ configurations corresponding to $\left|x_{1}\right\rangle$ and $(N-m)$ configurations corresponding to $\left|\alpha_{i}\right\rangle$.

\subsubsection*{Particles on a Line: without periodic boundary condition}
\begin{figure}[ht]
        \centering
        \includegraphics[width=6cm]{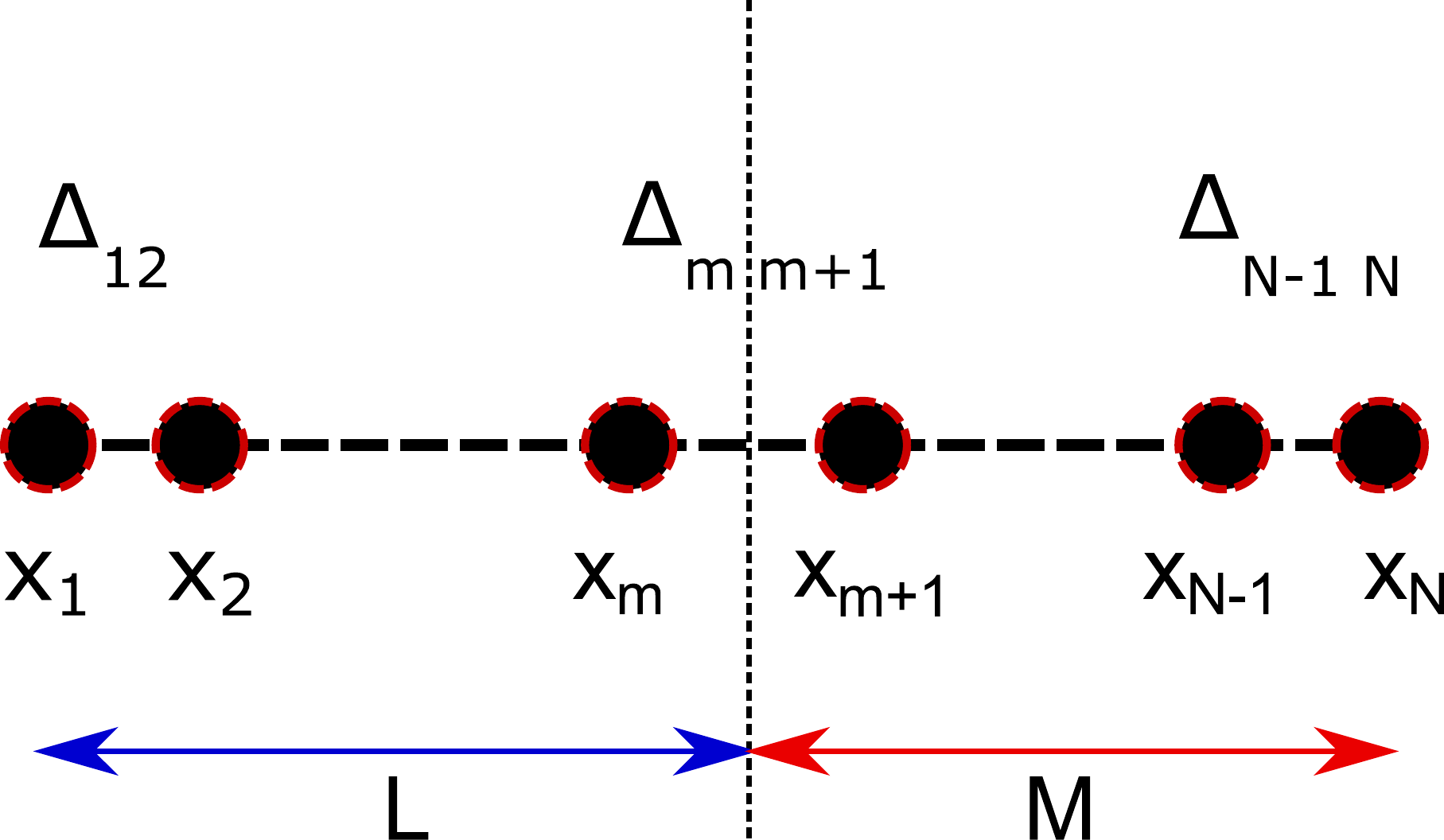}
        \caption{$N$ particles on a rod. $(x_1, x_{2},\cdots ,x_{N})$ denote the positions of these particles. $\Delta_{ij}$ denotes the separation between particles $i$ and $j$.  The first $m$ particles are traced out when we compute the reduced density matrix. }
        \label{fig:rod}
        \end{figure}
This section considers $N$ particles on a rod. The configuration is illustrated in Figure \ref{fig:rod}. The set-up is almost identical to the previous calculation but a periodic boundary condition is not imposed here. Hence, there is no constraint on these $\Delta_{ij}=x_i-x_j$, and they can take values independently of each other. We define $\sum_{i,j=1}^{m-1}\Delta_{ij}=L$ and $\sum_{i,j=m+1}^{N}\Delta_{ij}=M.$ Also $\Delta_{m-1\,m}$ can either take a value of $a$ or 0. The wavefunction can be written as
\be
\left|\Psi \right\rangle =\mathcal{N}\sum_{L,M}\Big(\left|x_{1},L\right\rangle \otimes \left|x_{M}=x_{1}+L,m\right\rangle + \left|x_1,L \right\rangle \otimes \left| x_m=x_1+L+a,M\right\rangle\Big).
\ee
The reduced density matrix is
\begin{align}
\begin{split}
\rho_{A}=&\Tr_{1\cdots m} \left|\psi\right\rangle \left\langle \psi \right| \\
=&\mathcal{N}^2\sum_{M,M', L=0}^{L=(m-1)a} \left\langle x_{1},L | x_{1}, L'\right\rangle \\ 
&\times \Big[ \left| x_{m}=x_1+L,M\right\rangle \left\langle x_{m}=x_+L,M'\right| \\ 
&\qquad + \left| x_{m}=x_1+L+a,M \right\rangle \left\langle x_{m}=x_{1}+L+a,M' \right| \\
&\qquad + \left| x_{m}=x_1+L,M \right\rangle \left\langle x_{m}=x_1+L+a,M' \right| \\
&\qquad + \left| x_{m}=x_1+L+a,M \right\rangle \left\langle x_{m}=x_1+L,M'\right| \Big] \,.
\end{split}
\end{align}
After some simplifications, one obtains
\begin{align}
\begin{split}
\rho_{A}=\mathcal{N}^2\sum_{M,M', L=0}^{L=(m-1)a} \Gamma(L)^2&\Big[2\, \left|x_{m},M \right\rangle \left\langle x_{m},M' \right| \\
&+ \left|x_{m},M\right\rangle\left\langle x_{m}+a,M'\right| \\
&+ \left|x_{m},M\right\rangle\left\langle x_{m}-a,M'\right| \Big] \,,
\end{split}
\end{align}
where $\Gamma(L)^2$ denotes all possible linear combinations of states. Finally, the reduced density matrix after summing over $L$ and $L'$ is
\begin{align}
\begin{split}
\rho_{A}=\mathcal{N}^2 \Delta \sum_{M,M'}&\Big[2\,\left|x_{m},M\right\rangle \left\langle x_{m},M'\right| \\
&+ \left|x_{m},M\right\rangle \left\langle x_{m}+a,M' \right| \\
&+ \left|x_{m},M\right\rangle\left\langle x_{m}-a,M'\right| \Big] \,.
\end{split}
\end{align}
This is almost the same as the 2-particle case in previous section except only with the extra factor of $\Delta$ which counts the degeneracy of the states.  From this the computation of $n=2$ R$\acute{\text{e}}$nyi entropy follows as before. 

One of the most important message from this calculation is that only the particles right next to the cut is responsible for the entanglement, as is testified by the form of the reduced density matrix. Therefore, the entanglement cannot exceed the dimension of the Hilbert space at site $m$, and therefore we can interpret this strictly as an ``area law'' in the 1 dimensional chain of particles. In this case the potential between particles is a rigid well where perturbation is not possible, and it resembles a gapped 1d system. The linear combination of states that represents the true ground state as a result of tunnelling, at the end, surprisingly did not lead to more entanglement beyond the area law even as such a linear combination resulting from tunnelling is taken into account.

\section{Tunnelling and entanglement entropy} \label{sec1}

To complete our discussion of quantum mechanical caricature of instantons, let us now make slightly less approximations, so that notions of instantons feature more obviously and prominently. 
We return to the 2-particle system, one at $x_1$ and the other at $x_2$. We consider an interaction potential that depends only  $x_1-x_2$. 
We will now investigate the situation where the potential has multiple minima, and so there is tunnelling between two square wells through a potential wall with a finite width. The exponential tail in the wavefunction is precisely equivalent to what we call the action of instantons, which we will discuss in closer detail in \cite{me} to make a detailed comparison with the corresponding field theoretic studies.  Here we want to investigate the entropy dependence on the width of the wall, and justify the approximations taken in the previous sections, up to this small exponentially suppressed correction. The main goal of this section is to mimic the situation of false vacuum decay \cite{Inst1} and get as much intuition as possible for the effect of tunnelling on the entanglement, particularly, how the exponentially suppressed tail first features in the time dependence.  
\subsection{Set-up}
Consider two square wells each of width $(b-a)$ are separated by a barrier of width $2a$. We would like to study states very close to the ground states. As is well known, there are two almost degenerate states, with the first excited state split from the ground state by exponential factors of the width of the potential barrier. The true ground state has even parity and the first excited state has odd parity, as we discuss in detail below.

The two particle wave-function can be obtained, including non-perturbative corrections. The even parity eigenstate wavefunction of a double square well with a barrier height of $V_0$ in between is \cite{qm1}
\begin{align}
 =& A \sin{k_{+}(x_1-x_2-b)} &\qquad a &< x_1-x_2 < b & \quad &\text{region I} \nonumber \\
\left| \Psi \right\rangle_{\text{even}} =& B \cosh{\kappa_{+}(x_1-x_2)} & \qquad -a &< x_1-x_2 < a &\quad &\text{region II} \nonumber \\
=& -A \sin{k_{+}(x_1-x_2 +b)} & \qquad -b &< x_1-x_2 < -a &\quad &\text{region III} \,,
\label{eqn:wavefunc}
\end{align}
where $k_{+}$ and $\kappa_{+}$ are related to the energy $E$ by
\be \label{eq:energy}
E = \frac{\hbar^2 k_{+}^2}{2m} = V_0 - \frac{\hbar^2 \kappa_{+}^2}{2m}  \,.
\ee
The odd parity wavefunction is
\begin{align}
=& A \sin{k_{-}(x_1-x_2-b)} &\qquad a &< x_1-x_2 < b & \quad &\text{region I} \nonumber \\
\left| \Psi \right\rangle_{\text{odd}} =& B \sinh{\kappa_{-}(x_1-x_2)} & \qquad -a &< x_1-x_2 < a &\quad &\text{region II} \nonumber \\
=& A \sin{k_{-}(x_1-x_2 +b)} & \qquad -b &< x_1-x_2 < -a &\quad &\text{region III} \,,
\label{eqn:wavefunc_odd}
\end{align}
\begin{figure}[ht]
        \centering
        \includegraphics[width=4cm]{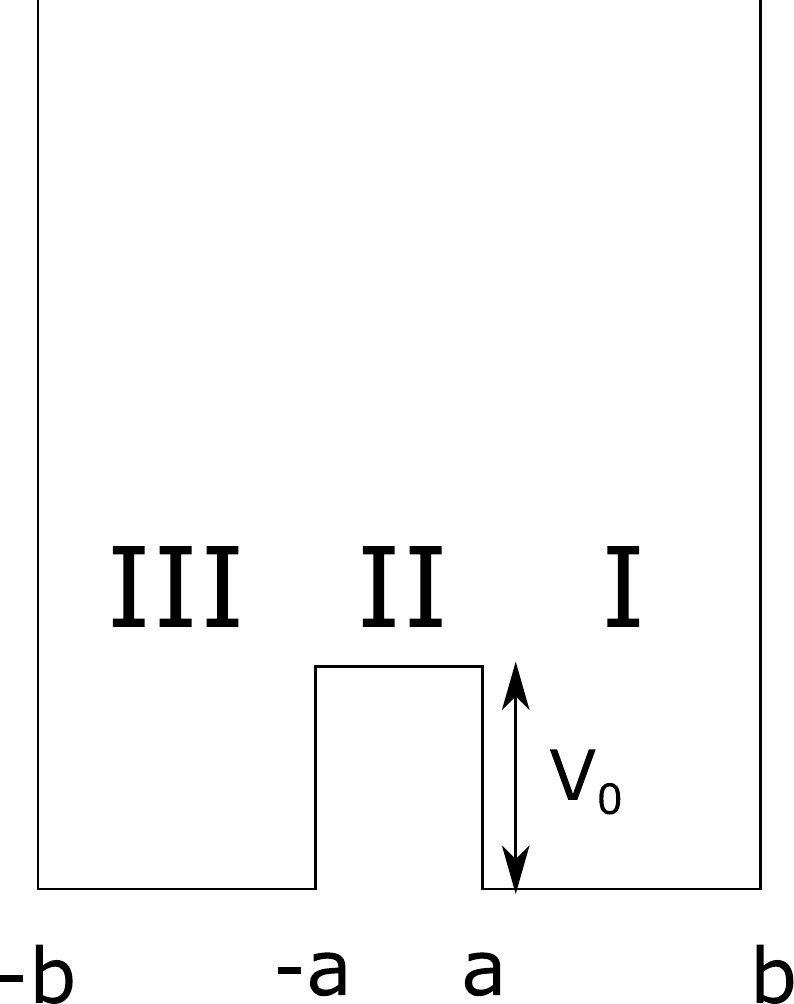}
        \caption{Square well.}
        \label{fig:sq_well}
\end{figure}
where $x_{1}$ and $x_{2}$ denote the positions of the two particles. The momenta $k_-, \kappa_-$ are related to the energy $E$ by the analogous relation (\ref{eq:energy}). This means that
\be \label{eq:potential}
V_0 = \frac{\hbar^2( k_+^2+ \kappa_+^2)}{2m}= \frac{\hbar^2( k_-^2+ \kappa_-^2)}{2m}
\ee
We will consider the limit where we can solve the double well problem entirely analytically.  There is a limit in which the eigenvalues can be obtained perturbatively without resorting to numerics.

Physically, we are expecting that the wavefunction in region I and III being very close to the ground state of the wells where they can be very well approximated by infinite wells, up to exponentially suppressed tails that leak into region II. Using this knowledge therefore,  
we  set $\delta k_{\pm}=  (\pi-\epsilon_{\pm }) $ where  $\delta= b-a.$, and we take $ \epsilon_{\pm} \ll 1$. The continuity conditions of the wavefunction at the walls lead to consistency conditions that constrain the eigenvalues $k_\pm$ and $\kappa_\pm$. For the even parity wavefunction, the consistency condition is
\be
k_{+}\cot(b-a)k_{+}=-\kappa_{+} \tanh a \kappa_{+},
\ee
and we get,
\be \label{var3}
k_{+}=\kappa_{+}\epsilon_{+}[1-e^{-2a\kappa_{+}}].
\ee
Similarly for the odd parity wavefunction, we get
\be
k_{-}=\kappa_{-}\epsilon_{-}[1+2e^{-2a\kappa_{-}}].
\ee
As $\epsilon_{\pm}$ are small so $k_{\pm}$ is very small compared to $\kappa_{\pm}$. Note however that $\kappa_\pm a$ are independent sets of dimensionless quantities, and to ensure that the tunnelling across the wells are exponentially suppressed, we will take $\kappa_{\pm} a$ to be large. 
Now using this we can solve for $\delta $ in terms of say $\epsilon_{-}.$
\be
\pi-\epsilon_{-}=\delta\, \kappa_{-}\epsilon_{-}[1-2e^{-2a \kappa_{-}}].
\ee
Upto the order we are interested in, we get
\be
\delta=\frac{\pi}{\kappa_{-}\epsilon_{-}}(1-\frac{\epsilon_{-}}{\pi}-2 e^{-2 a\kappa_{-}}).
\ee
For the later convenience, we can set
\be
\kappa_{-}=1.
\ee
Also we make the substitution
\be
a=\frac{\alpha}{\kappa_{-}}.
\ee
As promised,  $\alpha \gg 1$ but it's zeroth order in $\epsilon$. We have chosen to do a double expansion, in $1/\alpha$ and $\epsilon$ independently. To churn out physics out of a complicated expressions with many independent dimensionless quantities, some concessions are necessary.
Next we define
\be
\kappa_{+}=\kappa_{-}+\delta \kappa
\ee
and
\be
\epsilon_{+}=\epsilon_{-}+\delta\epsilon.
\ee
Using (\ref{eq:potential})
and
\be
\frac{\pi-\epsilon_{+}}{\pi-\epsilon_{-}}=\frac{\kappa_{+}\epsilon_{+}[1-e^{-2a\kappa_{+}}]}{\kappa_{-}\epsilon_{-}[1+2e^{-2a\kappa_{-}}]} \,,
\ee
we can solve for $\delta \kappa$ and $\delta \epsilon$. Finally we get
\be \label{var1}
\delta \kappa = \frac{4 \epsilon_{-}^3 \kappa}{\pi}e^{-2a\kappa_{-}} \,,
\ee
and 
\be\label{var2}
\delta \epsilon=4\epsilon_{-}e^{-2a\kappa_{-}}.
\ee
Now using (\ref{var1}),(\ref{var2}) and (\ref{var3}), we can write
\be
k_{+}=k_{-}+\delta k \,,
\ee
where
\be
\delta k= -\frac{4k \epsilon_-  e^{-2a\kappa_{-}}}{\pi}\, .
\ee

Normalising the wavefunctions and solving for the continuity conditions, we have
\begin{align}
A_{+}&=\frac{1}{\sqrt{L\delta}}(1-\frac{\epsilon_{-}}{2\pi}(1+4 e^{-2a\kappa_{-}}))\,,\\
A_{-}&=\frac{1}{\sqrt{L\delta}}(1-\frac{\epsilon_{-}}{2\pi})\,,\\
B_{-}&=-\frac{2\,\epsilon_{-}}{\sqrt{L\delta}}e^{-a\kappa_{-}}(1+e^{-2a\kappa_{-}})\,,\\
B_{+}&=\frac{2\epsilon_{-}e^{-a\kappa_{-}}}{\sqrt{L\delta}}(1+2e^{-2a\kappa_{-}}).
\end{align}
Typically the trace of the density matrix is divergent due to the space time volume divergence. It is regulated by introducing a cut-off $L$ to normalise the wavefunction. 
Finally, we compute the reduced density matrix when one particle has been traced out, and obtain from it the $n=2$ R$\acute{\text{e}}$nyi entropy
\be
S_{2}=-\ln (\tr \rho_{A}^2) \,.
\ee
Expanding in $\epsilon$ and $1/\alpha$ and keeping the leading terms
\begin{align} \label{eq:ren2a}
\begin{split}
L\tr \rho_{A}^2=&\bigg[\frac{(1-2 e^{-2 \alpha})}{12\pi\epsilon}\Big(153+20 \pi ^2-36\cos(\Delta E t)-(60+8\pi^2)\cos(2\Delta E t)\Big)\bigg]\\&+ \bigg[\Big\{\frac{(1+2\pi^2)e^{-2 \alpha}}{2 \pi ^2}-\frac{26 \left(6+\pi ^2\right) e^{-4 \alpha}}{3 \pi ^2}\Big\}-\\&\Big\{ \left(1+\frac{4 \left(3+\pi ^2\right) e^{-2 \alpha}}{3\,\pi ^2}-\frac{20 \left(3+\pi ^2\right) e^{-4 \alpha}}{3 \pi ^2}\right) \cos (\Delta E t)\Big\}\\&-\Big\{\left(\, e^{-2 \alpha}-\frac{10 \left(3+\pi ^2\right) e^{-4 \alpha}}{3 \pi ^2}\right)\cos (2 \Delta E t)\Big\}\bigg]+\mathcal{O}(\epsilon, \exp(-6\alpha)).
\end{split}
\end{align}

Note that there isn't any linear $t$ contribution. We will make more detailed comparison with a field theoretic instanton calculation in our another publication \cite{me}. Let us note however that the exponentially suppressed factors play exactly the same role as instantons, and that at small $t$ the time dependence starts at $t^2 \Delta E^2 \sim t^2 \exp(-4\alpha)$.
We note that in this calculation, we have taken a square double well problem which allowed us to solve for the wave-function exactly to compute entanglement entropy between the two particles. However, this problem is in fact the classic problem where the energy splitting $\delta \epsilon$ in (\ref{var2}) between the even and odd wave-functions can be approximated using the WKB approximation. A beautiful discussion of this subject can be found in \cite{Coleman}. We only quote here that
\be
\delta \epsilon = \hbar K \exp(-S_0/\hbar),
\ee
where $S_0$ is the action of the single instanton evaluated on-shell, and one can readily check that the instanton solution here does reproduce $\exp(-2 \alpha)$. We note therefore that its contribution to the entanglement entropy starts with $\exp(-2S_0/\hbar)$  which has the same action as a 2-instanton solution. There is no obvious reason a priori where single instanton solutions leading to linear $t$ terms do not contribute. In fact, it is a famous observation in quantum quenches in 1+1 d CFT that linear terms do contribute \cite{Cardy}, dictated by the so called ``light-cone'' effect at early times. 
We note however, that this result is highly reminiscent of our field theoretic result in \cite{me}, except that linear $t$ terms beginning with single instanton contributions to appear there.


\section{Free Bosons}

The computation of entanglement using the wavefunction method particularly for free field theory has been considered before \cite{CasiniFT}. What we would like to emphasize here, is that generically in a Lorentz invariant field theory, the local basis in configuration space is related to the mode operators $b$ and $b^\dag$ that defines the Lorentz invariant vacuum in the specific way due to causality. This linear combination is dictated uniquely by the vanishing of commutators for space-like separation.  See for example \cite{Weinberg} for a detailed discussion. Therefore, perhaps expected, the large amount of entanglement in configuration space follows from Lorentz invariance. This fact should already be apparent in the Unruh effect relevant particularly in the case of half-space entanglement. 
The crucial point is that if we construct local creation and annihilation operators -- i.e. creating bosons at site $x$ based on local field variables $a(x) = \phi(x)+ i \pi(x)$ and $a^\dag(x) = \phi(x) - i \pi(x)$, these local creation and annihilation operators are related to the set of creation and annihilation operators that defines a Lorentz invariant vacuum by a Bogoliubov transformation. i.e. In other words, engineering a Lorentz invariant vacuum such that all inertial observers agree that there are no excitations requires a lot of spatial entanglement. In the following, we will make these statements precise, by defining explicitly these local creation and annihilation basis, and recover the Bogoliubov transformation relating the operators $a$ and $b$. These Bogoliubov coefficients would then allow us to compute the entanglement entropy explicitly. 

We start by considering a free scalar field $\phi$ with mass $m$ and its conjugate momentum $\pi$,
\be
\binom{\phi}{\pi} = \int \frac{dk}{(2\pi)} \ \frac{1}{\sqrt{2E}} \left[ \binom{1}{-iE} b_k +  \binom{1}{iE} b^{\dagger}_{-k} \right] e^{ikx} \,,
\label{eqn:psi_mode}
\ee
where $E=\sqrt{k^2+m^2}$.

Let us consider a linear transformation which relates the set of operators $b_k$ to operators that are actually local
\be
\binom{\phi}{\pi} = \int \frac{dk}{(2\pi)} \ \binom{\phi_{k}}{\pi_{-k}} e^{ikx} \,.
\ee

\noindent Note that $\phi_k$ and $\pi_k$ satisfy the following commutation relation
\be
[ \phi_k,\pi_{k'} ] = i\delta_{k\,k'} \,.
\label{eqn:comm}
\ee
The two sets of operators are related by
\be
\binom{\phi_{k}}{\pi_{-k}} = \frac{1}{\sqrt{2E}}  
\begin{pmatrix}
        1 & 1  \\
        -iE & iE  \\
\end{pmatrix}
\binom{b_k}{b^{\dagger}_{-k}} \,.
\ee

\noindent One can express $(b_k,b_{-k}^{\dagger})$ in terms of $(\phi_{k},\pi_{k})$
\be
\binom{b_k}{b^{\dagger}_{-k}} = \frac{i}{\sqrt{2E}}
\begin{pmatrix}
        -iE & 1  \\
        -iE & -1  \\
\end{pmatrix}
\binom{\phi_{k}}{\pi_{-k}} \,.
\ee

The conditions above imply that $\phi^{\dagger}_{-k} = \phi_{k}$ and $\pi_{ -k}^{\dagger} = \pi_{k}$. These two conditions imply that $\phi_{k}$ must be the $\dagger$ of $\phi_{-k}$ and similarly for $\pi_{k}$ and $\pi_{-k}$. We still have the freedom to choose the relation between $\phi_k$ and $\pi_k$ which we will examine in the next section. Let us first define a vacuum $\left| 0 \right \rangle_m$ which is annihilated by the operator $b_k$
\be
b_k \left| 0 \right \rangle _m = 0 \,.
\ee
Hence, the new set of operators satisfy the equation

\be
\left( \sqrt{\frac{E}{2}} \, \phi_{k} + i \frac{1}{\sqrt{2E}} \, \pi_{-k} \right) \left| 0 \right \rangle_m = 0 \,.
\label{eqn:vac_eqn}
\ee

\subsection{Appropriate Bogoliubov transformation}
Here, we would like to make connection between the operators $b_k, b^\dag_k$ and the set of creation/annihilation operators that create/annihilate bosons locally. 
These local creation/annihilation operators $a(x), a^\dag(x)$ are defined as follows. 
\be
a(x) = \sqrt{\frac{m}{2}}\phi(x) + i \frac{1}{\sqrt{2m}} \pi(x) .
\ee
In which case, we have
\begin{align}
\phi_{k}&=\frac{1}{\sqrt{2m}} \left(a_k + a_{-k}^\dagger \right) \,, \label{eqn:phi_bog}\\
\pi_{-k} &=-i\sqrt{\frac{m}{2}} \left(a_k - a_{-k}^{\dagger} \right) \,. \label{eqn:pi_bog}
\end{align}

The linear map relating $a,a^\dag$ and $b, b^\dag$ is a Bogoliubov transformation. With this choice of transformation, the vacuum can be expressed in terms of a state constructed from the new set of operators
\begin{align}
\left| 0 \right \rangle_m &= \frac{1}{\gamma} e^{-\sum_{k} c_k a_k^{\dagger} a_{-k}^{\dagger}} \left| 0 \right \rangle \\
&= \frac{1}{\gamma} \prod_{k} \left( 1 - c_k a_k^{\dagger} a_{-k}^{\dagger} + O\left( (a_k^{\dagger})^2,(a_{-k}^{\dagger})^2 \right)\right) \left| 0 \right \rangle \,,
\label{eqn:vac_exp}
\end{align}
where $\gamma$ is the normalisation factor, $\left| 0 \right \rangle$ is the vacuum defined by $a_k \left| 0 \right \rangle = 0$ and the coefficients $c_k$ are fixed by equation (\ref{eqn:vac_eqn}) and it depends on our choice of Bogoliubov transformation. 

With (\ref{eqn:phi_bog}) and (\ref{eqn:pi_bog}), the coefficient $c_k$ in (\ref{eqn:vac_exp}) is

\be
c_k = \frac{E-m}{E+m} \,.
\ee

Let us now consider a system with finite number of sites. The inverse Fourier transform of the operator $a_k$ is
\be
a_k = \sum_{N} a_N e^{-ikN} \,,
\ee
where $N$ is the site label. The spatial annihilation and creation operators $a_N$ and $a_{N'}^\dagger$ satisfy the usual commutation relations
\be
[ a_N , a_{N'}^\dagger ] = \delta_{N,N'} \,.
\ee

The vacuum with respect to the operator $b_k$ can now be written as
\be \label{sup}
\left| 0 \right \rangle_m = \frac{1}{\gamma} \prod_{k} \left( 1 - \sum_{NL} c_k e^{ik(N-L)} a_N^{\dagger} a_{L}^{\dagger} \right) \left| 0 \right \rangle \,.
\ee

To make contact with our previous calculations -- what we have been taught by the quantum mechanical calculation is that the presence of multiple vacuua ensues that the wavefunction is generally a linear combinations of peaks over individual minimum. The true ground state after tunneling is taken into account would be the even combination of these peaks.
Now in a scalar field theory, a wavefunction peaking over a minimum basically controls the expectation value of the scalar field. This situation is reminiscent of superconductors. Quasi-particle excitations can be described using Bogoliubov transformations to rotate to suitable basis from some reference basis operators. 
Now we can put these intuition together to describe the true even-ground state wavefunctions of a field theory with two minima. It is basically a linear combination of two wavefunctions, each related to the local site creation and annihilation operators by a suitable bogoliubov transformation, whose coefficients are controlled to leading order in the perturbation around these respective minima, the effective masses.

Therefore, we inspect the entanglement entropy of  a state $\left| \Psi \right \rangle$ which is constructed from a general superposition of two vacua ($\alpha \left| 0 \right \rangle _{m_1} +\beta  \left| 0 \right \rangle _{m_2}$) defined by two sets of operators of the form (\ref{eqn:psi_mode}) with mass $m_1$ and $m_2$, respectively. Using (\ref{sup}) the $\left| \Psi \right \rangle$ takes the following form, 
\begin{align}
\left| \Psi \right \rangle = \frac{1}{\sqrt{\cal N}} \left( \prod_{k} \left( 1 - \sum_{NL} c_k e^{ik(N-L)} a_N^{\dagger} a_{L}^{\dagger} \right) \left| 0 \right \rangle + a_2 \prod_{k} \left( 1 - \sum_{NL} \tilde{c}_k e^{ik(N-L)} a_N^{\dagger} a_{L}^{\dagger} \right) \left| 0 \right \rangle \right) \,,
\end{align}
where $\cal N$ is a overall normalisation constant, $a_2$ is the relative weight factor, $c_k = \frac{E_1-m_1}{E_1+m_1}$ and $\tilde{c}_k=\frac{E_2-m_2}{E_2+m_2}$.
Assuming $m_1,m_2 \gg \frac{1}{\tilde{L}}$, where $\tilde{L}$ is the lattice spacing, the state can be simplified to
\begin{align}
\left| \Psi \right \rangle &= \frac{1}{\sqrt{\cal N}} \left( \left( 1 - \sum_{k} \sum_{NL} c_k e^{ik(N-L)} a_N^{\dagger} a_{L}^{\dagger} \right) \left| 0 \right \rangle + \left( a_2 - \sum_{k} \sum_{NL} a_2 \tilde{c}_k e^{ik(N-L)} a_N^{\dagger} a_{L}^{\dagger} \right) \left| 0 \right \rangle \right) \nonumber \\
&= \frac{1}{\sqrt{\cal N}} \left( 1+a_2 - \sum_{k} \sum_{NL} C_k f^k_{NL} a_N^{\dagger} a_{L}^{\dagger} \right) \left| 0 \right \rangle \,,
\end{align}
where $C_k = c_k + a_2\tilde{c}_k$ and $f^k_{NL}=e^{ik(N-L)}$. With this simplification, the normalization factors take the following form
\begin{align}
{\cal N} &= |1+a_2|^2 + \sum_{kk'} \sum_{NL} \left(f^k_{NL}f^{*k'}_{NL}+f^k_{NL}f^{*k'}_{LN} \right)C_k C^*_{k'} \\
&= |1+a_2|^2 + \kappa_0 + \kappa_1 + \kappa_2 \,
\end{align}
where
\begin{align}
\kappa_0 &= \sum_{kk'} \sum_{\bar{n}\bar{l}} C^*_{k'} C_k \left( f^{*k'}_{\bar{n}\bar{l}} f^k_{\bar{n}\bar{l}} + f^{*k'}_{\bar{l}\bar{n}} f^k_{\bar{n}\bar{l}} \right) \,, \label{eqn:sign_1} \\
\kappa_1 &= \sum_{kk'} \sum_{\bar{n}ll'} C^*_{k'} C_k \tilde{f}^{*k'}_{l'\bar{n}} \tilde{f}^k_{\bar{n}l} \, \\
\kappa_2 &= \sum_{kk'} \sum_{nl} C^*_{k'} C_k \left( f^{*k'}_{nl} f^k_{nl} + f^{*k'}_{ln} f^k_{nl} \right) \,. \label{eqn:sign_2}
\end{align}

To compute the entanglement entropy, the configuration space is divided into two regions $A$ and $\bar{A}$, where the sites in each region are labelled by small letters ($n$) and small letters with bar ($\bar{n}$), respectively. The state $\left| \Psi \right \rangle$ is
\begin{align}
\left| \Psi \right \rangle =& \frac{1}{\sqrt{\cal N}}\Big( (1+a_2) - \sum_{k} \sum_{nl} C_k f^k_{nl} a_n^{\dagger} a_{l}^{\dagger} - \sum_{k} \sum_{\bar{n} \bar{l}} C_k f^k_{\bar{n}\bar{l}} a_{\bar{n}}^{\dagger} a_{\bar{l}}^{\dagger}\nonumber  \\ &- \sum_{k} \sum_{\bar{n}l} C_k \tilde{f}^k_{\bar{n}l} a_{\bar{n}}^{\dagger} a_{l}^{\dagger} \Big) \left| 0 \right \rangle \,,
\end{align}
where $\tilde{f}^k_{\bar{n}l} = f^k_{\bar{n}l} + f^k_{l\bar{n}}$.

A reduced density matrix of $\rho_{A}$ is constructed by tracing out the degrees of freedom in region $\bar{A}$
\begin{align}
\rho_{A} &= \Tr_{\bar{A}} \left( \left| \Psi \right \rangle \left\langle \Psi \right| \right) \\
&= \frac{1}{\cal N} \left[ \left( (1+a_2) - \sum_{k} \sum_{nl} C_k f^k_{nl} a_n^{\dagger} a_{l}^{\dagger} \right) \left| 0 \right\rangle_{AA} \left\langle 0 \right| \left( (1+a_2^*) - \sum_{k'} \sum_{n'l'} C^*_{k'} f^{*k'}_{n'l'} a_{n'} a_{l'} \right) \right] {}_{\bar{A}} \left\langle 0 | 0 \right\rangle_{\bar{A}} \nonumber \\
& \ + \left| 0 \right\rangle_{AA}\left\langle 0 \right|
\left[ {}_{\bar{A}}\left\langle 0 \right| \left( \sum_{k'} \sum_{\bar{n}'\bar{l}'} C^*_{k'} f^{*k'}_{\bar{l}'\bar{n}'} a_{\bar{l}'} a_{\bar{n}'} \right) \left( \sum_{k} \sum_{\bar{n}\bar{l}} C_k f^k_{\bar{n}\bar{l}} a_{\bar{n}}^{\dagger} a_{\bar{l}}^{\dagger} \right) \left| 0 \right\rangle_{\bar{A}} \right] \nonumber\\
& \ + a_{l}^{\dagger} \left| 0 \right\rangle_{AA}\left\langle 0 \right| a_{l'} \left[ {}_{\bar{A}}\left\langle 0 \right| \left( \sum_{k'} \sum_{\bar{n}' l'} C^*_{k'} \tilde{f}^{*k'}_{l' \bar{n}'} a_{\bar{n}'} \right) \left( \sum_{k} \sum_{\bar{n} l} C_k \tilde{f}^k_{\bar{n} l} a_{\bar{n}}^{\dagger} \right) \left| 0 \right\rangle_{\bar{A}} \right] \,.
\end{align}

The reduced density matrix $\rho_{A}$ is then
\be
\rho_A = \frac{1}{\cal N} \left[ (|1+a_2|^2 +\kappa_0) \left| 0 \right\rangle \left\langle 0 \right| - (1+a_2^*)\left| 2 \right\rangle \left\langle 0 \right| - (1+a_2) \left| 0 \right\rangle \left\langle 2 \right| +  \left| 1 \right\rangle \left\langle 1 \right| +  \left| 2 \right\rangle \left\langle 2 \right| \right]\,,
\ee
where the subscript label $A$ for the vacuum in region $A$ is dropped, 
\begin{align}
\left| 2 \right\rangle &= \sum_{k} \sum_{nl} C_k f^k_{nl} a_n^{\dagger} a_{l}^{\dagger} \left| 0 \right\rangle \,, \\
\left| 1 \right\rangle \left\langle 1 \right| &= \sum_{kk'} \sum_{\bar{n}ll'} C^*_{k'} C_k \tilde{f}^{*k'}_{l'\bar{n}} \tilde{f}^k_{\bar{n}l} a_{l}^{\dagger} \left| 0 \right\rangle \left\langle 0 \right| a_{l'} \,.
\end{align}

Let us consider the second R$\acute{\text{e}}$nyi entropy $S_2$. The square of the reduced density matrix is 
\begin{align}
\rho_A^2 =& \frac{1}{{\cal N}^2}\left( (|1+a_2|^2 +\kappa_0)^2 + |1+a_2|^2\kappa_2 \right) \left| 0 \right\rangle \left\langle 0 \right| - \left( |1+a_2|^2(|1+a_2|^2 +\kappa_0) + (1+a_2)\kappa_2 \right)\left| 0 \right\rangle \left\langle 2 \right| \nonumber \\
&- \left( |1+a_2|^2(|1+a_2|^2 +\kappa_0) + (1+a_2^*)\kappa_2 \right)\left| 2 \right\rangle \left\langle 0 \right| + \tilde{\kappa} \left| 1 \right\rangle \left\langle 1 \right| + \left( |1+a_2|^2 + \kappa_2 \right)\left| 2 \right\rangle \left\langle 2 \right| \,. \label{eqn:rho_2}
\end{align}
The coefficients $\tilde{\kappa}$ is given by
\begin{align}
\tilde{\kappa} \left| 1 \right\rangle \left\langle 1 \right|&=  \left| 1 \right\rangle \left\langle 1 | 1 \right\rangle \left\langle 1 \right| \\
&= \left( \sum_{kk'} \sum_{\bar{n}lp} C^*_{k'} C_k \tilde{f}^{*k'}_{p\bar{n}} \tilde{f}^k_{\bar{n}l} a_{l}^{\dagger} \left| 0 \right\rangle \left\langle 0 \right| a_{p} \right) \left( \sum_{k''k'''} \sum_{\bar{n}'l'q} C^*_{k'''} C_{k''} \tilde{f}^{*k'''}_{l'\bar{n}'} \tilde{f}^{k''}_{\bar{n}'q} a_{q}^{\dagger} \left| 0 \right\rangle \left\langle 0 \right| a_{l'} \right) \nonumber \\
&= \sum_{kk'k''k'''} \sum_{\bar{n}\bar{n}'ll'p} C^*_{k'} C_k C^*_{k'''} C_{k''} \tilde{f}^{*k'}_{p\bar{n}} \tilde{f}^k_{\bar{n}l} \tilde{f}^{*k'''}_{l'\bar{n}'} \tilde{f}^{k''}_{\bar{n}'p} a_{l}^{\dagger} \left| 0 \right\rangle \left\langle 0 \right| a_{l'} \,,
\end{align}
$S_2$ is computed by taking the trace of (\ref{eqn:rho_2})
\begin{align}
S_2 &= - \ln \Tr \rho_A^2 \nonumber \\
&= - \ln \left( \frac{1}{{\cal N}^2} \left(|1+a_2|^4+ 2|1+a_2|^2\kappa_0 + 2|1+a_2|^2\kappa_2 + {\cal{O}}(c_k^4) \right) \right)\,, \nonumber \\
&= - \ln \left( 1 - 2\frac{\kappa_1}{|1+a_2|^2} + {\cal{O}}(c_k^4) \right)
\end{align}
where we have used ${\cal N}= |1+a_2|^2+\kappa_0 + \kappa_1 + \kappa_2$ and kept only the terms up to ${\cal{O}}(c_k^2)$ for consistency. $S_2$ contains only contributions from the links between region $A$ and $\bar{A}$ coming from $\kappa_1$, and there is no entanglement when correlation between region $A$ and $\bar{A}$ is turned off. 

\subsection{A discussion}

\begin{figure}[ht]
	\centering
	\includegraphics[width=14cm]{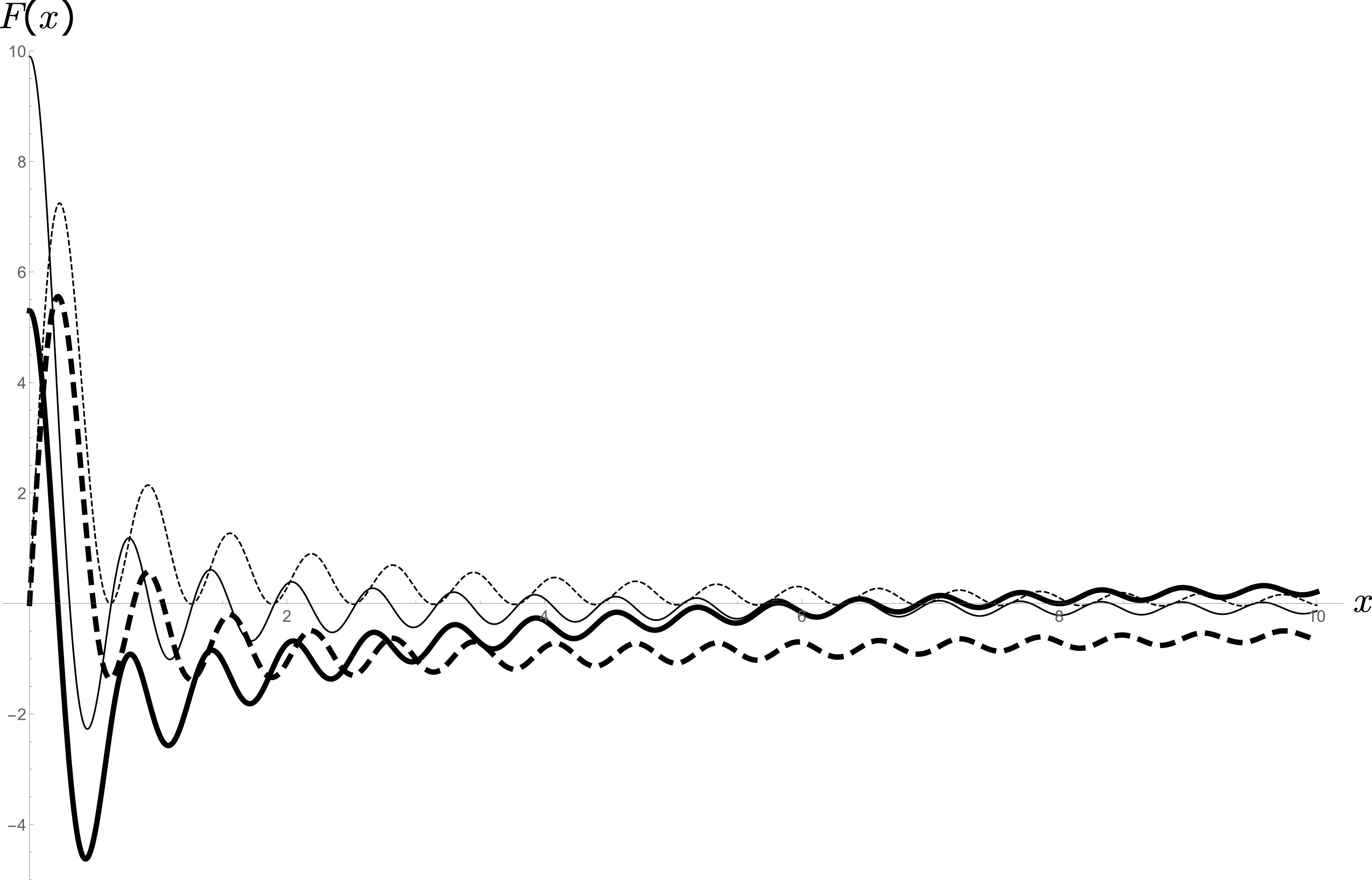}
	\caption{The thinner solid and dashed curves represent the real and imaginary parts of the integral up to the first order and the thicker curves represents the integral up to the second order, respectively. The amplitude of all curves decreases quickly with distance, and the long range contribution averages to zero due to the oscillatory behaviour. $m=1$, $\tilde{c}=0.1$ and $\tilde{L}=0.1$ are used in this plot.}
	\label{fig:integ_com}
\end{figure}

The coefficients $C_k=c_k + \tilde{c}_k$ control the range of the interaction. Recall that in this bosonic system
\be
c_k = \frac{E-m}{E+m} \,.
\ee
We expand the energy $E$ in powers of ($\frac{m}{k}$) and keep upto the second order. 
\be
\frac{E-m}{E+m} = 1 - \frac{2m}{k} + O\left((\frac{m}{k})^2 \right) \,.
\ee
The factor $\sum_k c_k f^k_{nl}$ can be converted to an integral by taking a continuum limit with an IR cut-off of the order of the mass $m$ and the inverse lattice spacing $\frac{1}{\tilde{L}}$ as the UV cut-off, respectively.
\begin{align}
F_{nl}=\sum_k c_k f^k_{nl} &\sim \int_{\tilde{c}m}^{\frac{1}{\tilde{L}}} \, dk \left( \frac{E-m}{E+m} \right) e^{ik(n-l)a} \\
&\approx F(x) \approx \int_{\tilde{c}m}^{\frac{1}{\tilde{L}}} \, dk \, \left( \frac{E-m}{E+m} \right) e^{ikx} \\
&\approx \int_{\tilde{c}m}^{\frac{1}{\tilde{L}}} \, dk \, \left( 1 - \frac{2m}{k} \right) e^{ikx} \,.
\end{align}
where $\tilde{c}$ is a constant. The integral is evaluated numerically and is shown in Figure \ref{fig:integ_com}.

This shows that the interaction between sites is short range-- the function decays exponentially with distances. Only the sites near the boundary contribute to the entanglement entropy. Hence, an area law is expected in this model \footnote{Interested readers are referred to the following references \cite{sq,sq1} and the references within for computation of entanglement entropy for squeezed states in various other situations.}.

\section{Discussions}
In this note, we attempt to study the physics of non-perturbative, or more specifically, instanton corrections to the entanglement entropy in a general many body system motivated by field theory. We pursue a strategy that focuses on the wavefunction itself. To set the stage, we look at the toy model of a many particle system where the two-particle potential has multiple minima. We apply the method of writing down explicit wavefunctions that describe the true ground states as in the case of field theory \cite{me}. The trick used here is the Bogoliubov transformation. Its utility is demonstrated in the simple case of free bosons in 1+1 dimensions, where the minima are massive. As a first test, we demonstrate that the area law is expected to hold. 
Our method attempts to attack the physics of the problem directly, which gives a very clear picture to the nature of the entanglement. In field theory, it can be understood as arising from pair creations of particles across a region and its complement dictated in a way demanded by Lorentz symmetry. Even after the effects of tunnelling is included, where the true ground state is approximated by a linear combination of vacua corresponding to peaks over different minima with generically different effective masses, we confirm that the entanglement entropy of these local theories still respects the area law in the presence of non-perturbative effects. 
We would like to explore how these methods can be applied also to gauge theories. That should provide a more direct insight to the nature of quantum entanglement in the ground states of gauge theories, and a handle to the reduced density matrix and its spectra.  

\section*{Acknowledgements}
Authors would like to thank Horacio Casini for valuable comments. AB and LYH would like to acknowledge support by the Thousand Young Talents Program, and Fudan University. SNL would like to acknowledge the support by Fudan University. AB, PHCL and LYH would like thank the organiser of the ``YKIS 2016: Quantum Matter, Spacetime and Information" conference held at YITP, Kyoto in June 13-17 where a part of this work was completed. PHCL thanks the hospitality of Fudan University and he is supported by JSPS postdoctoral fellowship for overseas researchers. Finally, AB, PHCL and LYH would like to thank Charles Melby-Thompson for various stimulating discussions and the collaboration on a paper \cite{me} related to this topic. 
\appendix
\section{Appendix}
\subsection{False vacuum tunnelling}
Let us go back to the 2-particle system and consider a configuration where the two particles are at the minimum $x=x_1-x_2=0$ at $t=0$. The particles will tunnel back and forth from $x=0$ to the other minimum $x=a$
\be
\left| \Psi_k(t,x_1,x_2) \right\rangle = \frac{\mathcal{N}}{2}\sum_{x_1,x_2} e^{i k X} \left[  |+\rangle   + e^{i\Delta t}  \left| -\right\rangle   \right] \,,\qquad |\pm\rangle = \left| x=0 \right\rangle  \pm\left| x=a \right\rangle,
\ee
where $\Delta$ is the energy difference between the ground state and the first excited state. The reduced density matrix of this state is
\be
\rho_1 = \mathcal{N}^2\sum_{x_1} \left( \left|x_1 \right\rangle \left\langle x_1 \right| -i\frac{\sin{\Delta t}}{2} e^{i \frac{k}{M} m_1 a} \left| x_1 \right \rangle \left\langle x_1+a \right| + i\frac{\sin{\Delta t}}{2} e^{- i \frac{k}{M} m_1 a} \left|x_1 \right\rangle \left\langle x_1-a \right| \right) \,.
\ee
The only difference from (\ref{eqn:red_den_2par}) is the additional sine factor in the off-diagonal terms. The contribution to $\Tr \rho_1^n$ from $m$ hops follows almost from previous consideration but a slight modification and is
\be
f_n(m)= \mathcal{N}^{2n}\frac{n!}{m!m!(n-2m)!} \left(\frac{\sin{\Delta t}}{2}\right)^{2m} \, .
\ee
Then the total contribution is
\be
\Tr \rho_{1}^{n}= \frac{1}{\mathcal{N}^2} \sum_{m=0}^{\left|\rfrac{n}{2}\right|} f_n(m) =\mathcal{N}^{2(n-1)} {}_2F_1 \Big(\frac{1}{2}-\frac{n}{2},-\frac{n}{2},1,\sin^2 \Delta t \Big) \,,
\ee
where ${}_2F_1$ is the hypergeometric function. A lower bound of the entanglement entropy is given by the second R$\acute{\text{e}}$nyi entropy $S_2$ as
\be
S_2=\mathcal{N}^{4} \left( 1 + \frac{\sin^2 \Delta t }{2} \right) \,.
\ee
Note that the leading time dependence in the ($\Delta t \ll 1$) limit is proportional to  $\Delta^2 t^2$. This suggest that the contribution is subleading compared to the usual instanton result in a field theory. 

We note that it is actually possible to have a linear time dependence in the R$\acute{\text{e}}$nyi entropy. Had we considered the following wave-function: 
\be
|\Psi\rangle = \frac{\mathcal{ N}}{\sqrt{2}}(A |+\rangle + B e^{i\Delta t} |-\rangle), \qquad |A|^2 +\ |B|^2 =1,
\ee
with generic complex numbers $A$ and $B$. The final result for the R$\acute{\text{e}}$nyi
entropy is given by
\be
S_n = \frac{\ln\tr(\rho^n_A)}{1-n}\,\qquad \tr(\rho_A^n) =  \mathcal{N}^{2(n-1)}{}_2F_1 \Big(\frac{1}{2}-\frac{n}{2},-\frac{n}{2},1,\sigma_+ \sigma_- \Big) ,
\ee
where we define
\be
\sigma_{+} = |A|^2 - |B|^2 \pm 2 i A \bar{B} \sin \Delta t \,\qquad \sigma_{-} = \bar{\sigma}_+.
\ee
A linear $t$ dependence appears as soon as $A$ and $B$ become complex numbers.
\subsection{Multi-particle}
This section presents an alternative method of computing the R$\acute{\text{e}}$nyi entropy of the multi-particle system. We start with the wavefunction for $N$ particles written in a similar form as in the two-particle case (\ref{eqn:2part}).
\begin{align}
\left| \Psi_k \right\rangle =& \frac{1}{\sqrt{2^{N}}} \sum_{\{x_i\}} \prod_{i=1}^{N-1} e^{ikX} \Big( \d(x_{i+1}-x_i) + \d(x_{i+1}-x_i-a) \Big) \nonumber \\
&\times \Big( \d(x_{N}-x_{1}) +\d(x_{N}-x_1-a) \Big) \left| x_1 \cdots x_{N} \right\rangle \,, \label{eqn:wavefunc}
\end{align}

The density matrix of the system is then
\begin{align}
\rho =& \left| \Psi_k \right\rangle \left\langle \Psi_k \right| \\
=& \frac{1}{2^{N}} \sum_{\{x_i\},\{y_i\}} \prod_{i=1}^{N-1} e^{ik(X-Y)} \nonumber \\ &\times \Big( \d(x_{i+1}-x_i) + \d(x_{i+1}-x_i-a) \Big) \Big( \d(y_{i+1}-y_i) + \d(y_{i+1}-y_i-a) \Big) \nonumber \\
&\times \Big( \d(x_{N}-x_{1}) +\d(x_{N}-x_1-a) \Big) \Big( \d(y_{2N}-y_{1}) +\d(y_{2N}-y_1-a) \Big)  \nonumber \\
&\times  \left| x_1 \cdots x_{N}\right\rangle \left\langle y_1 \cdots y_{N} \right| \,.
\end{align}

If one traces out the particles $(m+1,\cdots,N)$, the following reduced density matrix is obtained
\begin{align}
\rho_A =&  \frac{1}{2^{N}}\sum_{x_m,y_m}e^{i\frac{k}{M}m_m (x_m-y_m)} \sum_{x_1,y_1} e^{i\frac{k}{M}\sum_{j=1}^{m-1} m_j (x_1-y_1)} \\ \nonumber 
&\times \sum_{\tilde{k}_1=0}^{m-1} \d(x_m-x_1-\tilde{k}_1a)\sum_{s=\tilde{k}_1-1}^{\tilde{k}_1} \Big[\sum_{\tilde{N}_1=2}^{m-s}\sum_{\tilde{N}_2=\tilde{N}_1+1}^{m-s+1}\cdots\sum_{\tilde{N}_s=\tilde{N}_{s-1}+1}^{m-1}e^{i\frac{k}{M}\left(\sum_{i_1=\tilde{N}_1}^{m-1}m_{i_1}+\cdots+\sum_{i_s=\tilde{N}_s}^{m-1}m_{i_s}\right)a} \Big]\\　\nonumber 
&\times \sum_{\tilde{k}_2=0}^{m-1} \d(y_m-y_1-\tilde{k}_2a)\sum_{s=\tilde{k}_2-1}^{\tilde{k}_2} \Big[\sum_{\tilde{N}_1=2}^{m-s}\sum_{\tilde{N}_2=\tilde{N}_1+1}^{m-s+1}\cdots\sum_{\tilde{N}_s=\tilde{N}_{s-1}+1}^{m-1}e^{-i\frac{k}{M}\left(\sum_{i_1=\tilde{N}_1}^{m-1}m_{i_1}+\cdots+\sum_{i_s=\tilde{N}_s}^{m-1}m_{i_s}\right)a} \Big] \nonumber\\
&\times\sum_{x_{N}} \Big(\sum_{\tilde{k}=0}^{N-m} C^{N-m}_{\tilde{k}} \, \d(x_{N}-x_{m} -\tilde{k}a)
\d(x_{m}-y_m) \nonumber \\
&\qquad + \sum_{\tilde{k}=0}^{N-m-1} C^{N-m-1}_{\tilde{k}} \, \d(x_{N}-x_{m} -\tilde{k}a) \d(x_{m}-y_m-a) \nonumber \\
&\qquad + \sum_{\tilde{k}=1}^{N-m} C^{N-m}_{\tilde{k}} \, \d(x_{N}-x_{m} -\tilde{k}a) \d(y_{m}-x_m-a) \Big) \nonumber 
\end{align}
\begin{align}
&\times \Big( \d(x_{N}-x_1) + \d(x_{N}-x_1 -a) \Big)\Big( \d(x_{N}-y_1) + \d(x_{N}-y_1 -a) \Big) \nonumber \\
&\times \left| x_1 \cdots x_{m}\right\rangle \left\langle y_1 \cdots y_{m} \right| \,.
\end{align}

The $\d$-functions constrain the allowed value of $(\tilde{k}_1,\tilde{k}_2,\tilde{k}_3)$. A total of 8 different types of the components of the reduced density matrix is obtained.

\begin{align}
\frac{(N-m)+1}{2^{N}} \sum_{x_{N}}\left| x_{N}, \cdots ,x_{N}\right\rangle &\left\langle x_{N}, \cdots ,x_{N} \right| \label{eqn:wave1} \,,\\
 \frac{1}{2^{N}}  \sum_{x_{N}} e^{-i\frac{k}{M} \sum_{j=1}^{m-1} \tilde{m}_j a } \Big[\sum_{\tilde{N}_1=2}^{m-1} e^{i\frac{k}{M}\sum_{i_1=\tilde{N}_1}^{m-1}m_{i_1}a} \Big]　\left| x_{N}-a, \cdots ,x_{N} \right\rangle & \left\langle x_{N}, \cdots ,x_{N} \right| \label{eqn:wave2} \,,
 \end{align}
 \begin{align}
 \frac{1}{2^{N}}  \sum_{x_{N}} e^{i\frac{k}{M} \sum_{j=1}^{m-1} \tilde{m}_j a}
\Big[\sum_{\tilde{N}_1=2}^{m-1}e^{-i\frac{k}{M}\sum_{i_1=\tilde{N}_1}^{m-1}m_{i_1}a} \Big] \left| x_{N}, \cdots ,x_{N}\right\rangle &\left\langle x_{N}-a, \cdots ,x_{N} \right| \label{eqn:wave3} \,,\\
\frac{1}{2^{N}}  \sum_{x_{N}} \Big[\sum_{\tilde{N}_1=2}^{m-1}e^{i\frac{k}{M}\sum_{i_1=\tilde{N}_1}^{m-1}m_{i_1}a} \Big] \Big[\sum_{\tilde{N}_1=2}^{m-1}e^{-i\frac{k}{M}\sum_{i_1=\tilde{N}_1}^{m-1}m_{i_1}a} \Big] \left| x_{N}-a, \cdots ,x_{N}\right\rangle& \left\langle x_{N}-a, \cdots ,x_{N}\right| \label{eqn:wave4} \,,
\end{align}
 \begin{align}
\frac{1}{2^{N}} \sum_{x_{N}} e^{i\frac{k}{M} m_m a}  e^{i\frac{k}{M}\sum_{j=1}^{m-1} m_j a} \left| x_{N}, \cdots ,x_{N}\right\rangle &\left\langle x_{N}-a, \cdots ,x_{N}-a \right| \label{eqn:wave5} \,,\\
\frac{1}{2^{N}} \sum_{x_{N}}  e^{i\frac{k}{M} m_m a} \Big[\sum_{\tilde{N}_1=2}^{m-1} e^{i\frac{k}{M}\sum_{i_1=\tilde{N}_1}^{m-1}m_{i_1}a} \Big] \left| x_{N}-a, \cdots x_{N}\right\rangle& \left\langle x_{N}-a, \cdots ,x_{N}-a \right| \label{eqn:wave6} \,,\\
\frac{1}{2^{N}} \sum_{x_{N}} e^{-i\frac{k}{M} m_m a} e^{-i\frac{k}{M}\sum_{j=1}^{m-1} m_j a} \left| x_{N}-a \cdots x_{N}-a\right\rangle& \left\langle x_{N}, \cdots ,x_{N} \right| \label{eqn:wave7}\,, \\
\frac{1}{2^{N}} \sum_{x_{N}}  e^{-i\frac{k}{M} m_m a} \Big[\sum_{\tilde{N}_1=2}^{m-1}e^{-i\frac{k}{M}\sum_{i_1=\tilde{N}_1}^{m-1}m_{i_1}a} \Big] \left| x_{N}-a \cdots x_{N}-a\right\rangle &\left\langle x_{N}-a \cdots ,x_{N} \right| \label{eqn:wave8} \,.
\end{align}
The second R$\acute{\text{e}}$nyi entropy is
\begin{align}
S_2=& - \ln \left[ \left(\frac{1}{2^N}\right)^2 \left( N^2 + 2(m-1)^2 + 4(m-1) - 2N(m-1) + 2 \right) \right] \\
=&- \ln \left[ \left(\frac{1}{2^N}\right)^2 \left( N^2\left( \left(1-\frac{2}{N}\right) +2\left(\frac{m}{N}\right)^2 - 2\left(\frac{m}{N}\right) \right) \right) \right] \,,
\end{align}
which matches with (\ref{eqn:renyi2}).
\subsection{Double well tunnelling revisited}

In section \ref{sec1}, we discussed the 2 well potential in the limit that $k \ll \kappa$ and made some connections with instantons \cite{me}. Here, we would like to look for an alternative limit, in which $k \gg \kappa$. In this case the result takes a very similar form, but it does not have a simple semi-classical interpretation in terms of instantons. Nonetheless, we consider it for completeness. 

To take that limit, we follow the following procedure: \\
1.) We will first assume $b-a=\delta$ is very small so the well is very thin and set $a=1$. \\ 
2.) Next, set $k_{+}=\frac{1}{\delta} (\frac{\pi}{2}+\epsilon)$ and $k_{}=\frac{1}{\delta} (\frac{\pi}{2}+\epsilon)$, where $ \epsilon$ is a very small number.  \\
3.) Finally, we will carry out all the calculations by taking this double limit $\delta \rightarrow 0$ and $\epsilon \rightarrow 0$ such that the ratio of $\frac{\epsilon}{\delta}=\tilde \alpha$ remains finite.\\
Basically the consequence of these assumptions is that, the region II becomes very large and region III and I become very small. The wavefunction raises sharply in region III which reaches its maxima near $x_1-x_2=-a$, then decays very slowly in region II until it reaches its minima around the middle of region II. Then it raised again very slowly upto $x_1-x_2=a$ and finally decays very fast in region I. The length scale governing this in region I is $\frac{1}{k}$ which is smaller than the length scale $\frac{1}{\kappa}$ in region II. The consistency conditions coming from matching the wavefunction and its derivative at the boundary between region I and region II give the following condition \be
\kappa_{\pm}(1\mp 2 e^{-2 \kappa_{\pm} a})=k_{\pm}\epsilon \,.
\ee 
Then using the fact
\be k_{+}^2+\kappa_{+}^2=k_{-}^2+\kappa_{-}^2,
\ee
and assuming $\kappa_{+}=\kappa_{-}$, we have
\be
\frac{k_{+}}{k_{-}}^2=1-8 \epsilon^2 e^{-2\kappa a} \,.
\ee
Note that typically $\kappa$ is large but $\kappa a$ is very small. With our approximations, we can set 
\be  \label{expansion}
\kappa= \frac{\pi \tilde \alpha }{2}.
\ee
Next task is to determine the normalisation of the wavefunction $A$ and $B$. First we look at the even parity wavefunction. One equation comes from the continuity of the wavefunction at the boundary
\be \label{condition1}
\frac{B}{A}=2 e^{-\kappa a} (1-\frac{\epsilon^2}{2}) \,.
\ee
Another condition comes from the normalisation of the total density matrix
\be
\tr \rho=1 \,,
\ee
\be \label{condition2}
A^2 L \Big\{ [ (b-a)- \frac{\sin 2 k_{+}(b-a)}{2 k_+}]+\frac{B^2}{A^2}[a+\frac{\sinh 2\kappa a}{2 \kappa}]\Big\}=1 \,.
\ee
Solving (\ref{condition1}) and (\ref{condition2}), we get
\be
A^2=\frac{\kappa}{L} (1+e^{-4\kappa a}) \,,
\ee
and
\be
B^2= 4 \frac{\kappa }{L}e^{-2\kappa a}(1-\epsilon^2).
\ee
The wavefunction is translation invariant as it is a function of $x_1-x_2$. All the expressions are independent of $x_{1}$ after the integration over $x_2$. The integration over $x_{1}$ gives the length of the system which is typically divergent, so we impose an upper limit cut-off $L$ for the system size. For the odd parity wavefunction we also arrive at the same result for $A$ and $B$. Next we use (\ref{expansion}) and are left with only one parameter $\delta$. In all the subsequent calculations, we will expand around $\delta=0$ and extract the leading term.  
Now we want to compute the reduced density matrix for a wavefunction in a false vacuum and observe the effect of tunnelling to the entanglement entropy. Let us consider the following wavefunction
\be
\left|\psi\right\rangle=\alpha \left|+\right\rangle+ \beta e^{i \Delta E t } \left|-\right\rangle.
\ee
$\left|+\right\rangle$ and $\left|-\right\rangle$ respectively denote the even and odd parity wavefunctions, and $\alpha$ and $\beta$ can be complex in general but for simplicity we will set them to unity. $\Delta E$ is the energy splitting between the $\left|+\right\rangle$ and $\left|-\right\rangle$ states and is proportional to $e^{-\kappa a}$. As we want to observe the leading effect of energy splitting to the entropy, only the leading exponential terms is extracted. By virtue of our expansion scheme discussed in the previous section, we can keep track of the power of this exponential without expanding it. The first couple of terms is sufficient for our purpose. The leading contribution is
\be \label{leadingcontrib}
\tr \rho_{A}^2= \frac{2}{L\, \delta ^2}\left(\frac{5}{\pi  \tilde \alpha }-\frac{8}{3} e^{-E \pi }\pi^2 \tilde \alpha^2\right) +\cdots.
\ee
Note that the leading order is time independent. The time dependence typically comes from terms like $ e^{-2 \tilde \alpha  \pi } \tilde \alpha \cos(2\,\Delta E t) \sinh ( 2 \tilde \alpha \pi)$. As $\Delta E$ is itself proportional to $e^{- \kappa a}$, it is subleading compared to the terms in (\ref{leadingcontrib}). Substituting in $\tilde \alpha =\frac{2\kappa}{\pi}$ from (\ref{expansion}), one obtains (in $a=1$ convention)
\be 
\tr \rho_{A}^2=\frac{1}{L\, \delta ^2} \Big( \frac{5}{\kappa}- \frac{64}{3} e^{-4 \kappa }\kappa^2\Big).
\ee

\end{document}